\documentclass[conference]{IEEEtran}
\ifCLASSINFOpdf
\else
\fi
\usepackage[cmex10]{amsmath}
\ifCLASSOPTIONcompsoc
  \usepackage[caption=false,font=normalsize,labelfont=sf,textfont=sf]{subfig}
\else
  \usepackage[caption=false,font=footnotesize]{subfig}
\fi
\usepackage{fixltx2e}
\newcommand{\matr}[1]{\mathbf{#1}}
\DeclareMathOperator{\diag}{diag}

\DeclareMathOperator{\hermit}{H}
\DeclareMathOperator{\expec}{E}

\DeclareMathOperator*{\argmax}{arg\,max}
\DeclareMathOperator*{\argmin}{arg\,min}

\usepackage{bm}
\usepackage{stfloats}

\usepackage{pgfplots}
\usetikzlibrary{shapes}
\pgfplotsset{compat=newest}
\newlength\figureheight	
\newlength\figurewidth

\newcommand{\columnplot}{
\setlength\figureheight{0.232\textwidth}
\setlength\figurewidth{0.37\textwidth}
}	

\usetikzlibrary{external} 
\tikzexternaldisable

\usepackage[exponent-product=\cdot]{siunitx}
\DeclareSIUnit\dBm{dBm}
\DeclareSIUnit\dBi{dBi}

\begin{document}

%
\title{Measurement Results for Millimeter Wave pure LOS MIMO Channels}

\author{\IEEEauthorblockN{Tim H{\"a}lsig\IEEEauthorrefmark{1}, Darko Cvetkovski\IEEEauthorrefmark{2}, Eckhard Grass\IEEEauthorrefmark{2}, and Berthold Lankl\IEEEauthorrefmark{1}}
\IEEEauthorblockA{\IEEEauthorrefmark{1}Institute for Communications Engineering, Universit{\"a}t der Bundeswehr M{\"u}nchen, Germany\\
\IEEEauthorrefmark{2}Department of Computer Science, Humboldt-Universit{\"a}t zu Berlin, Germany\\
Email: tim.haelsig@unibw.de}
}


%


\maketitle

\begin{abstract}
In this paper we present measurement results for pure line-of-sight MIMO links operating in the millimeter wave range. We show that the estimated condition numbers and capacities of the measured channels are in good agreement with the theory for various transmission distances and antenna setups. Furthermore, the results show that orthogonal channel vectors can be observed if the spacing criterion is fulfilled, thus facilitating spatial multiplexing and achieving high spectral efficiencies even over fairly long distances. Spacings generating ill-conditioned channel matrices show on the other hand significantly reduced performance.
\end{abstract}


\let\thefootnote\relax\footnotetext{This work was supported in part by the German Research Foundation (DFG) in the framework of priority program SPP~1655 "Wireless Ultra High Data Rate Communication for Mobile Internet Access". We are indebted to IHP's system design department for providing some of the measurement equipment and assisting during the measurements.}

%
\IEEEpeerreviewmaketitle

\section{Introduction}
MIMO technology is nowadays the prevalent method when high spectral efficiencies and throughputs are needed in wireless communications systems. The maximal gain achievable by MIMO designs depends fundamentally on the channel characteristics, whereby parallel spatial streams, that yield the highest gain, can be transmitted if the channel vectors among the different receiving antennas are orthogonal to each other. 
For line-of-sight~(LOS) MIMO systems the channel characteristics yielding orthogonal vectors depend mostly on the antenna arrangement. With the advent of millimeter wave communications generating these optimal system designs, which depend on link distance, carrier frequency and number of spatial streams, is a viable option for different application scenarios, e.g., small cell wireless backhaul \cite{Cvetkovski2016}. Previous investigations have proven the concept in different frequency bands and for different system design ideas.

In \cite{Sheldon2008a,Sheldon2009} and other works by the same authors, a \SI{60}{\giga\hertz} LOS MIMO system with up to four spatially multiplexed streams and up to \SI{41}{\meter} link distance was implemented based on an analog channel separation network to recover the different streams. The works show that the streams can be well separated and that low BERs and high data rates can be achieved with this concept, but a sophisticated tuning of the analog network is required. Phase difference measurements, which give insight into the orthogonality of the channel vectors, for $2\times 2$ LOS MIMO systems operating at \SI{8}{\giga\hertz} and \SI{32}{\giga\hertz} over \SI{5.3}{\kilo\meter} and \SI{1.3}{\kilo\meter}, respectively, were presented in \cite{Bao2015}. The results agree well with LOS MIMO theory and show that long-term environmental influences, e.g., changing meteorological conditions, do not perturb the system excessively. A proof of concept for $2\times 2$ LOS MIMO over satellite was performed in \cite{Hofmann2016} at a carrier frequency of \SI{12}{\giga\hertz} and over a link distance of \SI{38200}{\kilo\meter}. The measured channel capacities match the predicted theoretical values and show that the orthogonal channel vector setup has a significantly improved performance as compared to a keyhole channel. 

In this paper we provide channel measurement results for a $2\times 2$ and $3\times 3$ LOS MIMO system setup operating at \SI{60}{\giga\hertz} that show the significant impact of the array geometry on the channel matrix and thereby potentially achievable throughput of LOS MIMO systems. We focus mainly on the condition number of the channel matrices as this gives a good insight into the orthogonality of the channel, the most important factor for spatial multiplexing. Different system arrangements, including different array spacings and offsets, are investigated for link distances up to \SI{60}{\meter}.

We denote transpose and conjugate transpose as $(\cdot)^T$ and $(\cdot)^H$. Boldface small letters, e.g., $\matr{x}$, are used for vectors while boldface capital letters, e.g., $\matr{X}$, are used for matrices. Furthermore, $\matr{I}_N$ denotes the $N\times N$ identity matrix.

\begin{figure*}[!t]
\centering
\subfloat[]{\def\antenna{%
   -- +(0mm,-2.5mm) -- +(0mm,0.75mm) -- +(1.31mm,2.5mm) -- +(-1.31mm,2.5mm) -- +(0mm,0.75mm)
}

\definecolor{mycolor1}{rgb}{0,0,0}%
\definecolor{mycolor2}{rgb}{1,1,1}%

\tikzset{>=latex}

\begin{tikzpicture}[font=\scriptsize]

\draw[color=mycolor1,fill=mycolor2,dashed] (0.35,0.15) -- (2.15,0.15);
\draw[color=mycolor1,fill=mycolor2,dashed] (2.15,2.05) -- (0.35,0.35);
\draw[color=mycolor1,fill=mycolor2,dashed] (0.35,2.25) -- (2.15,2.25);
\draw[color=mycolor1,fill=mycolor2,dashed] (0.35,2.05) -- (2.15,0.35);

\begin{scope}[xshift=0.75cm,yshift=0.825cm]
\draw[color=white,fill=white] (0,0) rectangle (1,0.65);
\end{scope}
\node[font=\normalsize,fill=white] at (1.25,1.15) {$\matr{H}_\text{LOS}$};


\draw[color=mycolor1,fill=mycolor2,thick] (0,0) \antenna;
\draw[fill=black] (0,0.9) circle (0.1mm);
\draw[fill=black] (0,1.0) circle (0.1mm);
\draw[fill=black] (0,1.1) circle (0.1mm);
\draw[color=mycolor1,fill=mycolor2,thick] (0,2) \antenna;

\draw[color=mycolor1,fill=mycolor2,thick] (-0.25,-0.25) -- (0,-0.25);
\draw[color=mycolor1,fill=mycolor2,thick] (-1.05,-0.5) rectangle (-0.25,0) node[pos=.5] {Tx $N$};
\draw[color=mycolor1,fill=mycolor2,thick,<-] (-1.05,-0.25) -- (-2.1,-0.25);

\begin{scope}[xshift=-0.4cm,yshift=1.25cm]
\draw[color=mycolor1,fill=mycolor2] (-1.4,-0.5) rectangle (-0.4,0) node[pos=.5] {Ref. Clk};
\end{scope}
\draw[color=mycolor1,fill=mycolor2,<-] (-0.65,0) |- (-0.8,1);
\draw[color=mycolor1,fill=mycolor2,<-] (-0.65,1.5) |- (-0.8,1);

\node[] at (-1.55,0) {$x_N(l_t)$};
\begin{scope}[yshift=2cm]
\node[] at (-1.55,0) {$x_1(l_t)$};
\end{scope}

\begin{scope}[yshift=2cm]
\draw[color=mycolor1,fill=mycolor2,thick] (-0.25,-0.25) -- (0,-0.25);
\draw[color=mycolor1,fill=mycolor2,thick] (-1.05,-0.5) rectangle (-0.25,0) node[pos=.5] {Tx $1$};
\draw[color=mycolor1,fill=mycolor2,thick,<-] (-1.05,-0.25) -- (-2.1,-0.25);
\end{scope}

\draw[color=mycolor1,fill=mycolor2,thick,align=center] (-2.8,-0.5) rectangle (-2.1,2) node[pos=.5] {AWG};
\node[align=center] at (-2.15,-0.75) {$f_s>\SI{10}{\giga Sa\per\second}$};

\draw[color=mycolor1,fill=mycolor2,thick] (2.5,0) \antenna;
\draw[fill=black] (2.5,0.9) circle (0.1mm);
\draw[fill=black] (2.5,1.0) circle (0.1mm);
\draw[fill=black] (2.5,1.1) circle (0.1mm);
\draw[color=mycolor1,fill=mycolor2,thick] (2.5,2) \antenna;

\draw[color=mycolor1,fill=mycolor2,thick] (2.5,-0.25) -- (2.75,-0.25);
\draw[color=mycolor1,fill=mycolor2,thick] (2.75,-0.5) rectangle (3.55,0) node[pos=.5] {Rx $M$};
\draw[color=mycolor1,fill=mycolor2,thick,<-] (4.6,-0.25) -- (3.55,-0.25);

\begin{scope}[xshift=4.7cm,yshift=1.25cm]
\draw[color=mycolor1,fill=mycolor2] (-1.4,-0.5) rectangle (-0.4,0) node[pos=.5] {Ref. Clk};
\end{scope}
\draw[color=mycolor1,fill=mycolor2,<-] (3.15,0) |- (3.3,1);
\draw[color=mycolor1,fill=mycolor2,<-] (3.15,1.5) |- (3.3,1);

\node[] at (4.05,0) {$y_M(l_t)$};
\begin{scope}[yshift=2cm]
\node[] at (4.05,0) {$y_1(l_t)$};
\end{scope}

\begin{scope}[yshift=2cm]
\draw[color=mycolor1,fill=mycolor2,thick] (2.5,-0.25) -- (2.75,-0.25);
\draw[color=mycolor1,fill=mycolor2,thick] (2.75,-0.5) rectangle (3.55,0) node[pos=.5] {Rx $1$};
\draw[color=mycolor1,fill=mycolor2,thick,<-] (4.6,-0.25) -- (3.55,-0.25);
\end{scope}

\draw[color=mycolor1,fill=mycolor2,thick,align=center] (4.6,-0.5) rectangle (5.4,2) node[pos=.5] {RTO(s)};
\node[] at (4.8,-0.75) {$f_s>\SI{5}{\giga Sa\per\second}$};

\node[font=\normalsize] at (1.25,-1.5) {\phantom{,}};

\end{tikzpicture}\label{fig:system_setup}}
\hfill
\subfloat[]{\input{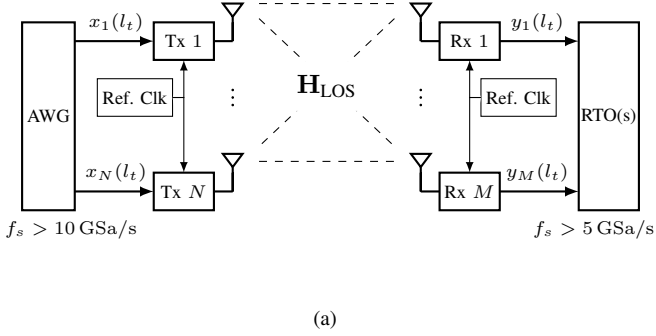}
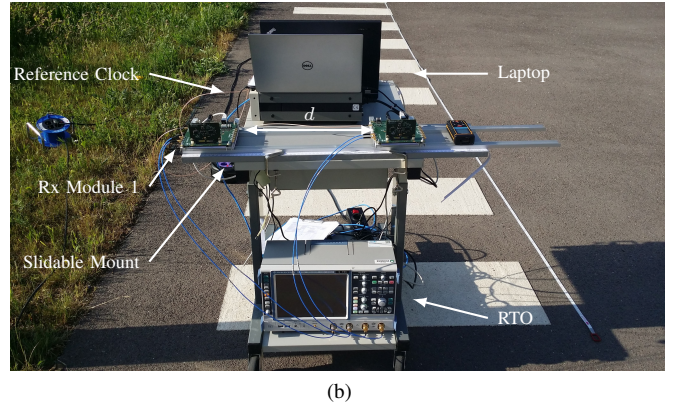\label{fig:foto_setup}}
\caption{Measurement system setup: \protect\subref{fig:system_setup}~Basic concept; \protect\subref{fig:foto_setup}~Receiver of one $2\times 2$ LOS MIMO setup, showing two \SI{60}{\giga\hertz} front-end modules, their shared reference clock generator, and an oscilloscope recording the received baseband signals.}
\label{fig:setup}
\end{figure*}

\section{Channel Model and Estimation}
We will first briefly describe the signal model and procedure used for extracting the channel coefficients from the recorded signals. Consider the discrete samples of the baseband signal received on the $m$th antenna to be given by the expression \cite{Ghogho2006}
\begin{equation}
\matr{y}_m = \matr{\Omega}\matr{X}\matr{h}_m + \matr{n}_m \text{,}\label{eq:model}
\end{equation}
which assumes that there is one shared normalized frequency offset $\Delta\omega$ across all transmit-receive antenna pairs, and where the vector $\matr{y}_m=\left[y_m(1)\phantom{,}\cdots\phantom{,}y_m(L_t)\right]^T$ collects the received signal of one channel training (or estimation) block of length $L_t$. The noise is assumed to be temporally and spatially uncorrelated with $\matr{n}_m=\left[n_m(1)\phantom{,}\cdots\phantom{,}n_m(L_t)\right]^T\sim\mathcal{C}\mathcal{N}(0,\sigma^2\matr{I}_{L_t})$. Furthermore, the channel vector of antenna $m$ is given by $\matr{h}_{m}=\left[\matr{h}_{m1}^T\phantom{,}\cdots\phantom{,}\matr{h}_{mN}^T\right]^T$ where each entry of the vector is given by a finite tap channel response with $L_c$ significant taps as $\matr{h}_{mn}=\left[h_{mn}(1)\phantom{,}\cdots\phantom{,}h_{mn}(L_c)\phantom{,}\cdots\phantom{,}h_{mn}(L_t)\right]^T$, with $N$ and $M$ denoting the number of transmit and receive antennas, respectively. 
Finally, the frequency offset between transmitter and receiver for the received block is expressed in the diagonal matrix $\matr{\Omega}=\diag(e^{j\Delta\omega},e^{j2\Delta\omega},\ldots,e^{jL_t\Delta\omega})$ and the training signal block of all transmit antennas is given by $\matr{X}=\left[\matr{X}_1\phantom{,}\cdots\phantom{,}\matr{X}_N\right]$ with
\begin{align}
\matr{X}_n &=
\begin{bmatrix}
x_n(1) & x_n(0) & \cdots & x_n(-L_t+2)\\
x_n(2) & x_n(1) & \cdots & x_n(-L_t+3)\\
\vdots & \vdots & \ddots & \vdots\\
x_n(L_t) & x_n(L_t-1) & \cdots & x_n(1)\\
\end{bmatrix}
\end{align}
which has a Toeplitz structure in order to represent the convolution of the training signal with the channel response.

For the estimation of the unknown channel vector and frequency offset, the Gaussian property of the noise is exploited. In the Gaussian noise case, the maximum likelihood estimator for the two values is given by
\begin{align}
\left(\hat{\matr{h}}_m,\Delta\hat{\omega}\right) &= \argmax_{\matr{h}_{m},\Delta\omega} P\left(\matr{y}_m|\matr{h}_{m},\Delta\omega\right) \\
 &= \argmin_{\matr{h}_m,\Delta\omega}\left\| \matr{y}_m-\matr{\Omega}\matr{X}\matr{h}_m\right\|^2 \text{.}
\end{align}
With orthogonal training sequences from different antennas, i.e., $\matr{X}\matr{X}^{\hermit}=\matr{I}_{L_t}$, and some matrix manipulations, see also \cite{Ghogho2006}, we get to the channel estimate with 
\begin{align}
\hat{\matr{h}}_{mn} &= \matr{X}_n^{\hermit}\matr{\Omega}^{\hermit}\matr{y}_m \label{eq:h_est}
\end{align}
which corresponds to correlating the received signal with the corresponding transmitted test signal after removing the estimated frequency offset. For estimation of the normalized frequency offset we use the estimator given by
\begin{equation}
\Delta\hat{\omega} = \frac{1}{MN}\sum_{\forall m,n}\frac{1}{(L-1) L_c}\sum_{l=1}^{L-1}\sum_{l_t=1}^{L_c}\frac{1}{L_t}\arg\left(\frac{c_{mn,l}(l_t)}{c_{mn,l+1}(l_t)}\right) \text{,}\label{eq:om_est}
\end{equation}
where $L\geq 2$ is the number of consecutive realizations of the training sequence over which the channel can be considered quasi-static, and $\matr{c}_{mn,l}=\matr{X}_n^{\hermit}\matr{y}_{m,l}=\left[c_{mn,l}(1)\phantom{,}\cdots\phantom{,}c_{mn,l}(L_t)\right]^T$ is the initial correlation output. Note that this estimator only uses the $L_c$ significant channel taps for the estimation of $\Delta\hat{\omega}$, the remaining entries up to $L_t$ will be used for noise power estimation later on.


\subsection{LOS MIMO and Performance Evaluation}
We are here solely interested in the pure LOS component of the channel, i.e., the entry of each channel impulse response estimate $\hat{\matr{h}}_{mn}$ with the highest magnitude. Thus, after using \eqref{eq:om_est} and \eqref{eq:h_est} consecutively to find the channel impulse response between antenna $n$ and $m$, we use $\hat{h}_{mn,\text{LOS}}=\max_{l_t}\hat{h}_{mn}(l_t)$ to get the LOS components, yielding the LOS channel matrix
\begin{equation}
\hat{\matr{H}}_\text{LOS} =
\begin{bmatrix}
\hat{h}_{11,\text{LOS}} & \cdots & \hat{h}_{1N,\text{LOS}}\\
\vdots & \ddots & \vdots \\
\hat{h}_{M1,\text{LOS}}& \cdots & \hat{h}_{MN,\text{LOS}} \\
\end{bmatrix} \text{,}
\end{equation}
where the entries are theoretically given by \cite{Halsig2015}
\begin{equation}
h_{mn,\text{LOS}} = a_{mn}\cdot\exp\left(-j2\pi\frac{r_{mn}}{\lambda}\right)
\end{equation}
with $\lambda$ being the wavelength of the carrier frequency, $r_{mn}$ being the distance between transmit antenna $n$ and receive antenna $m$, and $a_{mn}$ being an attenuation coefficient depending on the link distance and the gains of the used transmitter and receiver chains. Theoretically, these attenuation values should be very similar across the different paths in a LOS scenario. However, due to differences in the transceivers, e.g, amplifier gains and antenna patterns, these coefficients varied in our case in the order of 50\% across the different matrix entries. To check only the phase relations between the channel vectors we introduce the normalization
\begin{equation}
\left(\hat{\matr{H}}_\text{LOS,norm}\right)_{mn} = \frac{\hat{h}_{mn,\text{LOS}}}{\left|\hat{h}_{mn,\text{LOS}}\right|} \text{.} \label{eq:h_norm}
\end{equation}

We assess the performance of different setups by using the condition number of the channel matrices, which gives direct insight into the orthogonality of the channel matrix and is very sensitive even to small variations as will be shown later. It is defined by
\begin{equation}
\kappa = \frac{\lambda_{\max}\left({\matr{H}}_\text{LOS}\right)}{\lambda_{\min}\left({\matr{H}}_\text{LOS}\right)} \label{eq:cond}
\end{equation}
where $\lambda_{\min}(\cdot)$ and $\lambda_{\max}(\cdot)$ give the smallest and largest eigenvalue of a matrix, respectively, and we have orthogonal channel vectors if $\kappa=1$. Additionally, we provide some LOS MIMO capacity results, based on the well known equation
\begin{equation}
C = \log_2 \left(\det\left(\matr{I}_M+\rho\cdot\matr{H}_\text{LOS} \matr{H}_\text{LOS}^{\hermit}\right)\right) \label{eq:cap}
\end{equation}
with $\rho$ being the average signal-to-noise~ratio~(SNR), which we estimate using
\begin{equation}
\hat{\rho} = \frac{1}{MN}\sum_{m=1}^{M}\sum_{n=1}^{N}\frac{|\hat{h}_{mn,\text{LOS}}|^2}{ \hat{\sigma}_{mn}^2} \label{eq:SNRest}
\end{equation}
where the noise level is determined using
\begin{equation}
\hat{\sigma}_{mn}^2 = \frac{1}{L_t-L_c}\sum_{l_t=L_c+1}^{L_t}|\hat{h}_{mn}(l_t)|^2 \text{.}
\end{equation}
Note that the estimation procedures described here require that $L_t>L_c$, i.e., the training sequence should be longer than the impulse response of the channel.

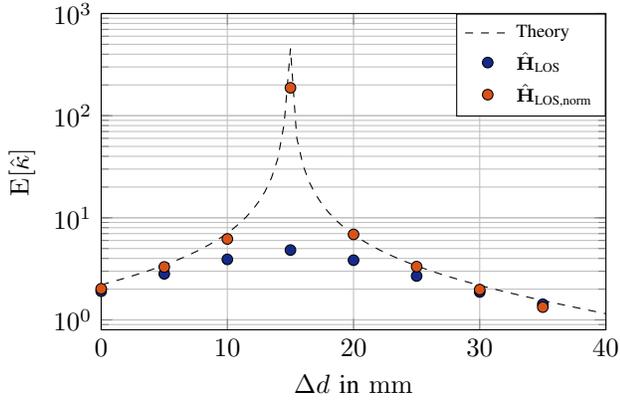
\begin{figure}[!t]
\centering
\columnplot
%
%
%
\definecolor{mycolor1}{rgb}{0.00000,0.44700,0.74100}%
\definecolor{mycolor2}{rgb}{0.85000,0.32500,0.09800}%
\definecolor{mycolor3}{rgb}{0.92900,0.69400,0.12500}%
\definecolor{usualBlue}{rgb}{0.0784313753247261,0.168627455830574,0.549019634723663}

\begin{tikzpicture}

\begin{axis}[%
width=\figurewidth,
height=\figureheight,
scale only axis,
xmin=0,
xmax=40,
xlabel={$\Delta d$ in \SI{}{\milli\meter}},
xmajorgrids,
ymode=log,
ymin=8*10^-1,
ymax=1000,
ylabel={$\expec [\hat{\kappa}]$},
yminorticks=true,
ymajorgrids,
yminorgrids,
legend style={at={(1.00,1.00)},legend columns=1,anchor=north east,draw=black,fill=white,legend cell align=left,font=\scriptsize}
]

\addplot [color=black,solid,dashed]
  table[row sep=crcr]{0	2.20795816540631\\
0.5	2.29435732698776\\
1	2.38659748333023\\
1.5	2.48533311478267\\
2	2.59131992021986\\
2.5	2.70543518794748\\
3	2.82870329232899\\
3.5	2.96232788672608\\
4	3.10773293844539\\
4.5	3.26661557495972\\
5	3.44101490755989\\
5.5	3.63340276719679\\
6	3.84680494783811\\
6.5	4.08496563464851\\
7	4.35257409487255\\
7.5	4.65558299025243\\
8	5.00166463380956\\
8.5	5.40088036353712\\
9	5.86668897542393\\
9.5	6.41751305960829\\
10	7.07925991046613\\
10.5	7.8895524467084\\
11	8.90519591485065\\
11.5	10.2161893123857\\
12	11.9741191874981\\
12.5	14.4557142490014\\
13	18.2255685589605\\
13.5	24.6420121854109\\
14	38.0078697857447\\
14.5	82.9884216962177\\
15	453.288839930078\\
15.5	60.740905893749\\
16	32.5438896480971\\
16.5	22.2197534814611\\
17	16.8631237196483\\
17.5	13.5829745081442\\
18	11.3671695555988\\
18.5	9.76942626297988\\
19	8.56236857016831\\
19.5	7.61796800584576\\
20	6.85862689624238\\
20.5	6.23457329375039\\
21	5.7124145381722\\
21.5	5.26891164041517\\
22	4.887392616924\\
22.5	4.55558584011861\\
23	4.26425809913242\\
23.5	4.00632900866975\\
24	3.77627809857145\\
24.5	3.56973762549395\\
25	3.38320660308049\\
25.5	3.21384594007202\\
26	3.05932905949306\\
26.5	2.91773122626709\\
27	2.78744636561945\\
27.5	2.66712372218793\\
28	2.55561904974569\\
28.5	2.45195658577292\\
29	2.35529912952361\\
29.5	2.26492427791366\\
30	2.18020538980445\\
30.5	2.1005962162744\\
31	2.02561839880175\\
31.5	1.95485122984608\\
32	1.88792321216163\\
32.5	1.82450505867009\\
33	1.76430385398523\\
33.5	1.7070581587238\\
34	1.65253388365242\\
34.5	1.60052079601689\\
35	1.55082954786343\\
35.5	1.5032891375618\\
36	1.45774473258247\\
36.5	1.41405579494871\\
37	1.37209446135815\\
37.5	1.33174413850412\\
38	1.29289828095384\\
38.5	1.25545932448125\\
39	1.21933775225902\\
39.5	1.18445127497865\\
40	1.15072410900733\\
};
\addlegendentry{Theory};

\addplot [color=black,only marks,mark=*,mark options={solid,scale=1,fill=usualBlue}]
  table[row sep=crcr]{0	1.90922167\\
5	2.824149063\\
10	3.914769895\\
15	4.831284401\\
20	3.843672655\\
25	2.689926399\\
30	1.877800652\\
35	1.422705493\\
};
\addlegendentry{$\hat{\matr{H}}_\text{LOS}$};

\addplot [color=black,only marks,mark=*,mark options={solid,fill=mycolor2}]
  table[row sep=crcr]{0	2.019302823\\
5	3.299899821\\
10	6.202447529\\
15	187.5244353\\
20	6.875662221\\
25	3.325387836\\
30	1.982093291\\
35	1.334258218\\
};
\addlegendentry{$\hat{\matr{H}}_\text{LOS,norm}$};

\end{axis}
\end{tikzpicture}%
\caption{Mean of the estimated condition numbers for a $2\times 2$ LOS MIMO setup with different spacing offsets $\Delta d$ on one of the transmitting modules for a fixed link range $R$. Measurements reveal the sensitivity with respect to that translation and show good agreement with the predicted theoretical values.}
\label{fig:xshift}
\end{figure}

\section{Measurement Setup}
The measurement setup consists of two subsystems at the transmitter and receiver, which will be briefly described in the next two subsections. The system setup is shown in Fig.~\ref{fig:setup}.

\subsection{Signal Generation, Recording and Control}
The first subsystem consists of the signal generation, processing and control devices. The baseband test signals for probing the channel were generated by an arbitrary~waveform~generator~(AWG) at the transmitter and sampled by a real-time~oscilloscope~(RTO) at the receiver with sampling rates $f_s$. They were each controlled by a laptop in order to provide a quick way to generate and save the used baseband waveforms, and allow for preliminary processing at the receiver. The laptops furthermore controlled the front-end settings as described further on. The test signals $x_n(l_t)$ supplied to the inputs of the front-ends covered a bandwidth of at least \SI{700}{\mega\hertz} to reflect the actual transmission schemes planned for this type of system. More specifically, orthogonal m-sequences of length $L_t=1023$ with good correlation properties and an oversampling factor of at least 8 were used at the different transmitters in order to probe the channel. The recorded signals are then processed as follows:
\begin{enumerate}
\item Use the $l$th received block $\matr{y}_{m,l}$ and $\matr{X}_n$ to get an initial correlation output $\matr{c}_{mn,l}$
\item Estimate $\Delta\hat{\omega}$ using \eqref{eq:om_est}, then estimate $\hat{\matr{h}}_{mn}$ using \eqref{eq:h_est}
\item Extract the LOS channel components to get $\hat{\matr{H}}_\text{LOS}$ and perform normalization according to \eqref{eq:h_norm}
\item Compute condition number estimate $\hat{\kappa}$ \eqref{eq:cond}, SNR estimate $\hat{\rho}$ \eqref{eq:SNRest} and channel capacity estimate $\hat{C}$ \eqref{eq:cap}
\end{enumerate}

For the $3\times 3$ scenario, see Fig.~\ref{fig:3by3}, the setup had to be slightly modified because six recording channels were necessary at the receiver (I and Q baseband for each receiver). Since most RTOs have at most four recording channels, we used two separate RTOs and coupled them by applying an external trigger signal and a common reference clock to assure reasonable synchronization when capturing the received signals. Nevertheless, an additional alignment of the different received signals based on the initial correlation output was needed in order to get coherent estimates.

\begin{figure}[!t]
\centering
\input{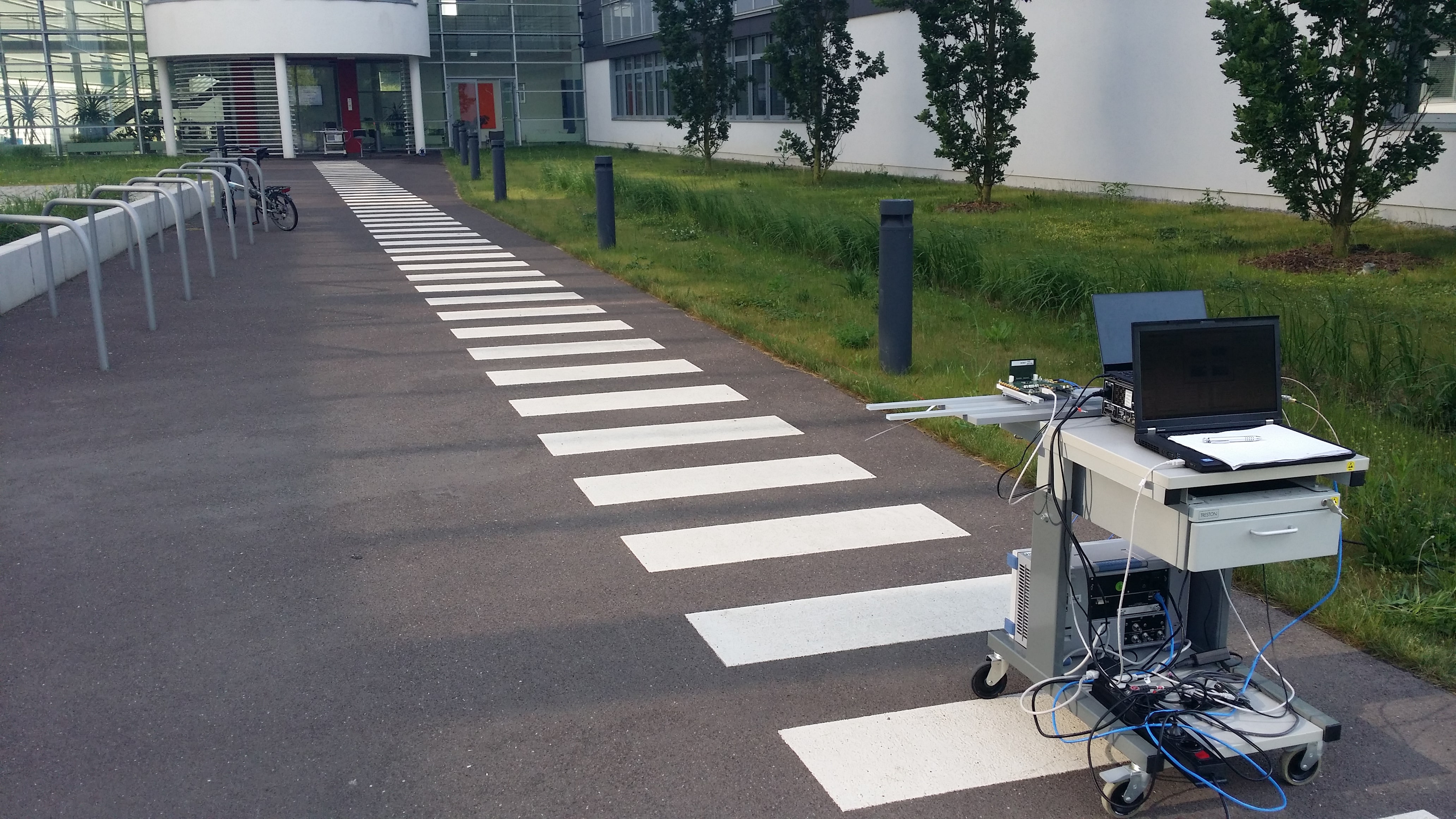}
\caption{Measurement scenario for $2\times 2$ MIMO along a LOS path for different transmission distances $R$.}
\label{fig:outside}
\end{figure}

\begin{figure}[!t]
\centering
\columnplot
\input{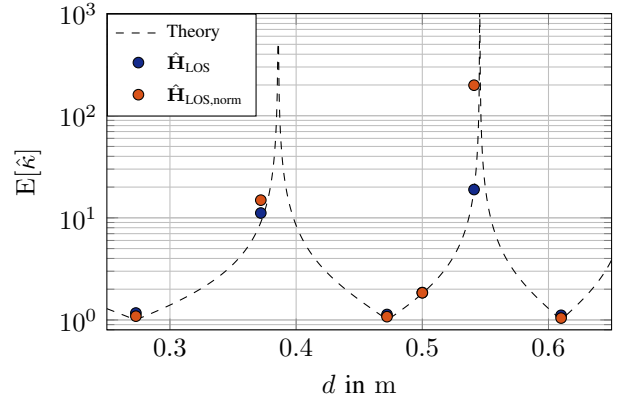}
\caption{Mean of the estimated condition numbers for a $2\times 2$ LOS MIMO setup with different module spacings for distance $R=\SI{30}{\meter}$, showing optimal and non-optimal spacings.}
\label{fig:sep30m}
\end{figure}

\begin{figure*}[!t]
\centering
\subfloat[]{\columnplot\input{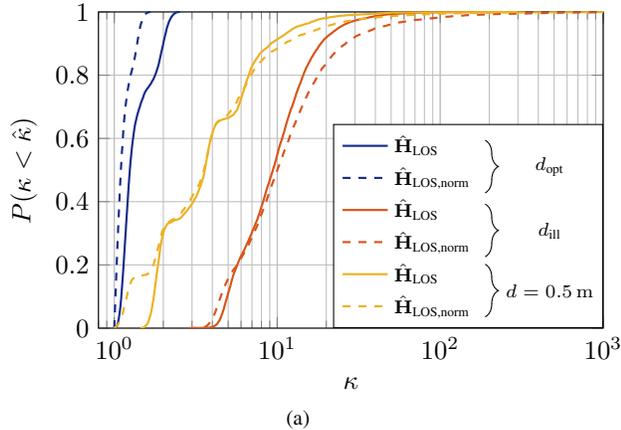}
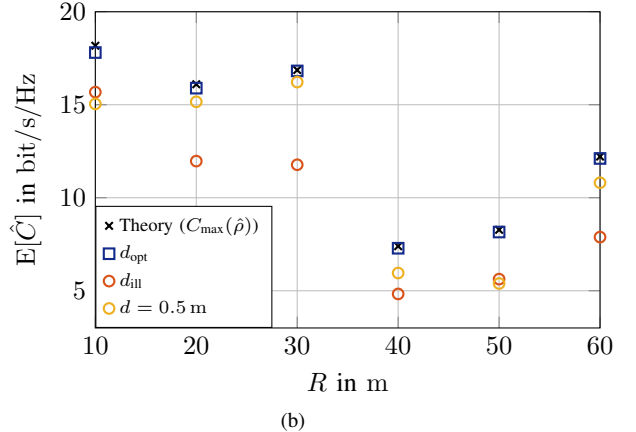\label{fig:cdf1}}
\hfill
\subfloat[]{\columnplot
%
%
%
\definecolor{mycolor1}{rgb}{0.00000,0.44700,0.74100}%
\definecolor{mycolor2}{rgb}{0.85000,0.32500,0.09800}%
\definecolor{mycolor3}{rgb}{0.92900,0.69400,0.12500}%
\definecolor{usualBlue}{rgb}{0.0784313753247261,0.168627455830574,0.549019634723663}

\begin{tikzpicture}

\begin{axis}[%
width=\figurewidth,
height=\figureheight,
scale only axis,
xmin=10,
xmax=60,
xmajorgrids,
xlabel={$R$ in \SI{}{\meter}},
ymin=3,
ymax=20,
ylabel={$\expec [\hat{C}]$ in \SI{}{bit\per\second\per\hertz}},
ymajorgrids,
legend style={at={(0.00,0.00)},legend columns=1,anchor=south west,draw=black,fill=white,legend cell align=left,font=\scriptsize}
]

\addplot [color=black,only marks,mark=x,mark options={solid},thick]
  table[row sep=crcr]{10	18.173327823171\\
20	16.1014948844983\\
30	16.8650831618434\\
40	7.39568407016513\\
50	8.27564132990945\\
60	12.2140774864815\\
};
\addlegendentry{Theory ($C_\text{max}(\hat{\rho})$)};

\addplot [color=usualBlue,only marks,mark=square,mark options={solid},thick]
  table[row sep=crcr]{10	17.8070440687678\\
20	15.8924933039358\\
30	16.8175688032309\\
40	7.28582759026104\\
50	8.1575372957257\\
60	12.1123744542222\\
};
\addlegendentry{$d_\text{opt}$};

\addplot [color=mycolor2,only marks,mark=o,mark options={solid},thick]
  table[row sep=crcr]{10	15.6788835609696\\
20	11.9702897843773\\
30	11.7741206626452\\
40	4.82954825101904\\
50	5.62566394140195\\
60	7.88532162630692\\
};
\addlegendentry{$d_\text{ill}$};

\addplot [color=mycolor3,only marks,mark=o,mark options={solid},thick]
  table[row sep=crcr]{10	15.0444944189939\\
20	15.1587354846323\\
30	16.2182747123805\\
40	5.95477508956762\\
50	5.39000725852908\\
60	10.8070979545221\\
};
\addlegendentry{$d=\SI{0.5}{\meter}$};

\end{axis}
\end{tikzpicture}%
\label{fig:cap1}}
\caption{Results of a $2\times 2$ LOS MIMO setup for different link distances $R$ and module spacings $d$: \protect\subref{fig:cdf1}~CDF of the estimated condition numbers of all distances; \protect\subref{fig:cap1}~Mean channel capacity based on unnormalized channel estimates $\hat{\matr{H}}_\text{LOS}$ and estimated SNR $\hat{\rho}$.}
\label{fig:collect_R}
\end{figure*}

\subsection{RF Front-End}
The second subsystem consists of the RF front-end and all devices related to it. We used the Hittite HMC600x integrated front-end transceivers with selectable carrier frequencies between \SI{57}{}-\SI{64}{\giga\hertz} facilitating an RF bandwidth of \SI{1.8}{\giga\hertz}. The transmitter can generate an equivalent isotropically radiated power of \SI{23}{\dBm} and has an in-package antenna with a fairly wide beam, generating a gain of \SI{7.5}{\dBi}. External reference clocks were shared among the modules on the transmitter and receiver side independently, which are used by the PLL of each module to generate the selected carrier frequency of \SI{60.48}{\giga\hertz}. This was done in order to simplify frequency offset estimation \cite{Halsig2016} and to yield the model in \eqref{eq:model}. Note that the difference between the Tx/Rx reference clocks upscaled by the PLLs causes the encountered frequency offset.

Due to the short wavelength, i.e., $\lambda=\SI{5}{\milli\meter}$, even small displacements of the modules in the millimeter range influence the channel matrix significantly. To allow for a precise control of the setup, the modules were thus fixed on a slidable mount with a displacement resolution of \SI{1}{\milli\meter}.

\section{Measurement Scenarios}
Most of the results presented in this section are analyzed with respect to different spacings $d$ between the modules, see Fig.~\ref{fig:setup}, as this parameter has the most significant impact on the performance of the $N\times M$ LOS MIMO link. There are numerous optimal, i.e., $\kappa=1$, spacings $d_\text{opt}$ and numerous ill-conditioned, i.e., $\kappa\to\infty$, spacings $d_\text{ill}$ for every scenario. The spacing criterion can be found, e.g., in \cite{Sheldon2009,Halsig2015}. Note that, except for the results in section~\ref{sec:disp_res}, we always used the same $d$ for the transmitter and receiver side. The typical snapshot of one channel recording was a few milliseconds, for which the channel was seen to be close to time-invariant with our setup.

\subsection{Displacement of one Transmit Module in x-Direction} \label{sec:disp_res}
In the first setup a link distance of $R=\SI{1.993}{\meter}$ was used and an initial module spacing of $d=\SI{0.18}{\meter}$ was set. Then, transmitter module 1 was offset by $\Delta d$ from the initial spacing in the same direction since this is the offset that the setup is most sensitive to. The mean of the estimated condition number $\expec [\hat{\kappa}]$, where we averaged over 3000 realizations, is shown in Fig.~\ref{fig:xshift}. The results agree well with the theoretical predicted ones, especially when only the phase relations are considered, i.e., $\hat{\matr{H}}_\text{LOS,norm}$. Notably, one ill-conditioned spacing is found in this setup as $d_\text{ill}=\SI{0.195}{\meter}$ at $\Delta d=\SI{15}{\milli\meter}$.

\begin{figure}[!t]
\centering
\columnplot
\input{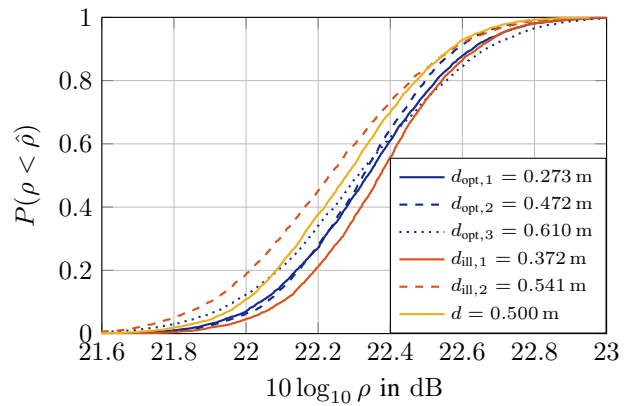}
\caption{CDF of the estimated SNR for a $2\times 2$ LOS MIMO link with $R=\SI{30}{\meter}$ at different module spacings.}
\label{fig:estSNR_pos}
\end{figure}

\subsection{Optimal and ill-conditioned Spacings for different Link Ranges}
For this setup we measured the channel at several link ranges $R=\SI{10}{\meter},\ldots,\SI{60}{\meter}$, see Fig.~\ref{fig:outside}, with their respective $d_\text{opt}$, $d_\text{ill}$ and the fixed spacing of $d=\SI{0.5}{\meter}$. As an example, the estimated condition numbers for $R=\SI{30}{\meter}$, averaged over 3000 realizations, are given in Fig.~\ref{fig:sep30m}. Again, the results are in good agreement with the theory and optimal channel conditions could be achieved, e.g., with the spacings $d_{\text{opt},1}=\SI{0.273}{\meter}$ and $d_{\text{opt},2}=\SI{0.472}{\meter}$.

\begin{figure*}[!t]
\centering
\subfloat[]{\columnplot\input{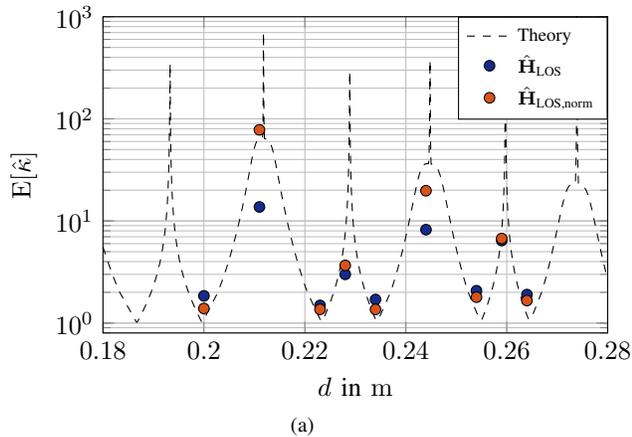}\label{fig:cond_3by3}}
\hfill
\subfloat[]{\columnplot\input{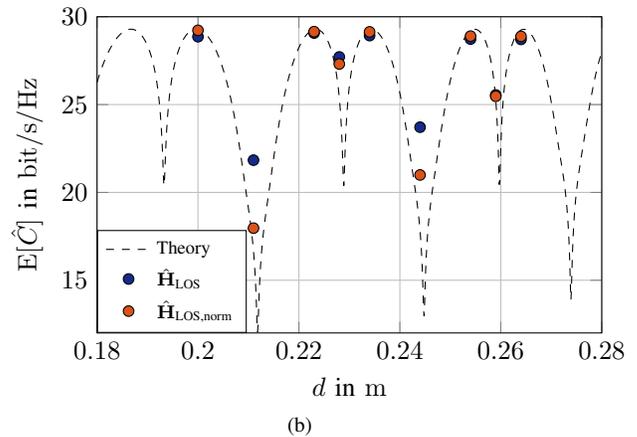}\label{fig:cap_3by3}}
\caption{Results of a $3\times 3$ LOS MIMO setup for a link distance of \SI{3}{\meter} and different module spacings: \protect\subref{fig:cond_3by3}~Mean of the estimated condition numbers; \protect\subref{fig:cap_3by3}~Mean of the estimated channel capacity, the estimated SNR is \SI{24.6}{\decibel}.}
\label{fig:3by3_results}
\end{figure*}

In Fig.~\ref{fig:collect_R} we have combined the results for all measurement scenarios and distances. The first plot shows the cumulative~distribution~function~(CDF) of the condition number for all distances classified into: optimal spacing ($d_\text{opt}$), ill-conditioned spacing ($d_\text{ill}$) and fixed spacing ($d=\SI{0.5}{\meter}$); where 800 realizations were used per spacing and distance. The results show that for the optimal spacings the condition number is generally low, for the ill-conditioned spacings it is significantly higher and for the fixed spacing it lies somewhere in between, thus being a good trade-off if the optimal positions cannot be achieved. The variance for each curve comes from different impairment effects, such as thermal noise, phase noise, gain differences and non-ideal link alignment. In the second plot we show the mean channel capacity of the different distances for the different spacing classes based on $\hat{\matr{H}}_\text{LOS}$ and $\hat{\rho}$ from \eqref{eq:SNRest}. The general trend is similar to the previous results. A key observation is that for the optimal spacings the theoretical maximum of the capacity is achieved almost exactly, i.e., orthogonal channel vectors were observed. This also shows that the condition number is the much more stringent performance measure in terms of showing the potential of the channel to support spatial multiplexing, since even condition numbers up to $\kappa=2$ can achieve capacities very close to the theoretical maximum. The variation of the channel capacity over the distances comes from the variation of the estimated SNR $\hat{\rho}$ and more specifically the variation of the received signal power, possibly due to fading effects or alignment errors when setting up the link.

The CDFs for the estimated SNRs of different measurement spacings at $R=\SI{30}{\meter}$ can be found in Fig.~\ref{fig:estSNR_pos}. While $\hat{\rho}$ varies notably with the link distance, there is only a small dependence on the spacing $d$ of the modules.

\subsection{Different Spacings for a Short-Range $3\times 3$ Link}
\begin{figure}[!t]
\centering
\input{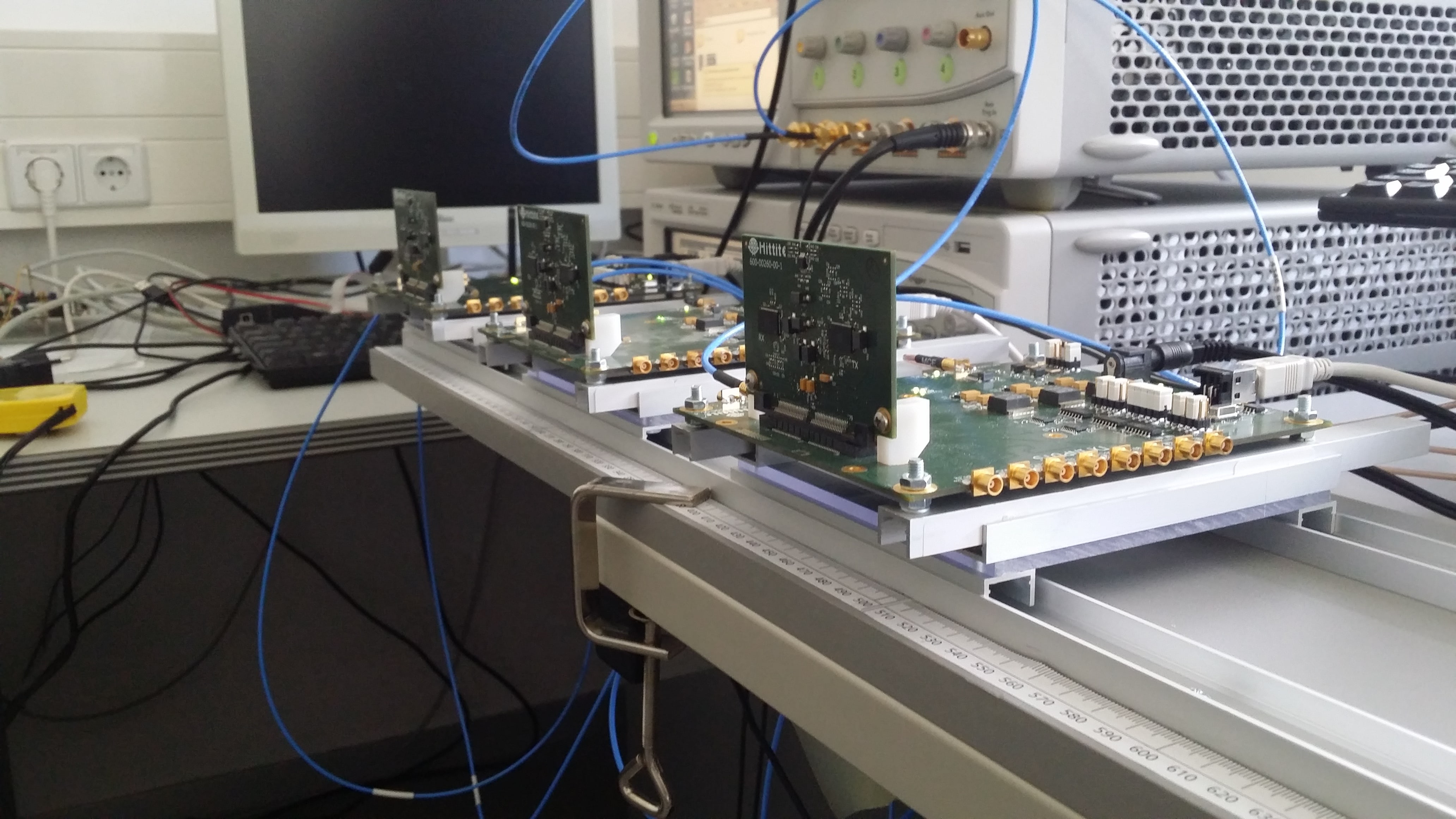}
\caption{Receiver of a $3\times 3$ LOS MIMO setup, showing three \SI{60}{\giga\hertz} modules, two oscilloscopes recording the received baseband signals, as well as the slidable mount for changing the spacing $d$ between the modules.}
\label{fig:3by3}
\end{figure}

Finally, we also present measurement results for a $3\times 3$ LOS MIMO link, see Fig.~\ref{fig:3by3}, with a link distance of $R=\SI{3}{\meter}$. Since the measurement setup is very bulky for the three antenna case, we only present measurements over a fairly short distance in a laboratory environment. We again set antenna spacings resulting in optimal and ill-conditioned channel matrices. Due to reduced memory of the measurement equipment we averaged only over 400 realizations per spacing. Results for the condition number and the capacity, for which the estimated mean SNR was $10\log_{10}\hat{\rho}=\SI{24.6}{\decibel}$, are presented in Fig.~\ref{fig:3by3_results}. As for the $2\times 2$ case we used equidistant and equal spacing on the transmitter and receiver side.

The results for the estimated condition number follow the trend predicted by theory, but the optimal and ill-conditioned cases are less distinct compared to the two antenna case. This is to be expected since the condition number is increasingly sensitive to small offsets and misalignments when the number of antennas in the system increases. The capacity results for the optimal points match very closely with the theory, i.e., orthogonal channel vectors were created. Although the results for the ill-conditioned spacings are notably lower than the optimal ones, they are significantly higher than what is predicted from theory. This can be explained by mainly two factors. First, the notch where the channel matrix is ill-conditioned gets narrower with shorter link distances and is thus much harder to set, since even small deviations improve the condition number markedly. Secondly, both phase and thermal noise can lead to an overestimation by adding favorable noise contributions to the channel entries which can excite eigenmodes that the true channel would not excite, as for example mentioned in \cite{Hofmann2016,Kyritsi2002}. Finally, the ill-conditioned spacings performance could also have been improved compared to the theoretical one by unresolvable multipath due to the short link distance and cluttered lab environment.

\section{Conclusion}
In this paper we have presented measurement results for a $2\times 2$ and $3\times 3$ pure LOS MIMO link at \SI{60}{\giga\hertz} for different link ranges. We have shown that the variation of the condition number with respect to different offsets and spacings of the observed channels is in good agreement the theoretically predicted one. Furthermore, the results show that very low condition numbers can be achieved if the optimal spacing criterion can be fulfilled. Subsequently, the estimated channel capacity results reveal that spatial multiplexing is viable in these setups. To that end we achieved estimated maximum capacities of \SI{16.9}{bit\per\second\per\hertz} and \SI{29.2}{bit\per\second\per\hertz} at estimated measurement SNRs of \SI{24.3}{\decibel} and \SI{24.6}{\decibel} for the $2\times 2$ and $3\times 3$ setup, respectively. The results also show that the ill-conditioned spacings have significantly increased condition numbers and reduced capacities for all of the measured scenarios and should thus be avoided if maximum system throughput is desired. 

In general, our results as well as the results in \cite{Sheldon2008a,Bao2015,Hofmann2016,Knopp2007} show that spherical wave propagation, which makes spatial multiplexing in pure LOS channels possible, can be observed over various frequency bands and for fairly long link distances. With respect to our measurement setup, higher antenna gain and/or higher output power in combination with appropriate module spacing could support spatial multiplexing at even longer link distances.

\bibliographystyle{IEEEtran}
\bibliography{references}
%

\end{document}